\documentclass[12pt]{article}
\usepackage{verbatim}
\usepackage{amsfonts}
\usepackage{graphics}
\usepackage{amsmath}
\usepackage{times}
\usepackage{appendix}
\usepackage{graphicx}
\usepackage{color}
\usepackage{float}
\usepackage{enumerate}
\usepackage{fancyhdr,latexsym,amsmath,amsfonts,amssymb,amsbsy,amsthm,url}
\usepackage[margin=0.5in,footskip=0.25in]{geometry}
\usepackage{graphics,graphicx,epsfig}
\usepackage{breqn}
\usepackage{subcaption}

\usepackage[english]{babel}
\usepackage[dvipsnames]{xcolor}
\usepackage{tikz}

\numberwithin{equation}{section}

\fancypagestyle{usual}{
	\lhead[\thepage]{}
	\chead[{\it \shortauthors}]{\sc \shorttitle}
	\rhead[]{\thepage}
	\lfoot[]{}
	\cfoot[]{}
	\rfoot[]{}
	
}

\makeatletter

\renewcommand{\section}{
	\@startsection
	{section}
	{1}
	{0pt}
	{1.1\baselineskip}
	{0.2\baselineskip}
	{\sc \centering}
}

\renewcommand{\subsection}{
	\@startsection
	{subsection}
	{1}
	{0pt}
	{1.1\baselineskip}
	{0.2\baselineskip}
	{\sc \centering}
}

\renewcommand{\subsubsection}{
	\@startsection
	{subsubsection}
	{1}
	{0pt}
	{1.1\baselineskip}
	{0.2\baselineskip}
	{\sc \centering}
}

\makeatother

\begin{document}
\title{\large\sc Studying the age of onset and detection of Chronic Myeloid Leukemia using a three-stage stochastic model}
\author{\sc{Suryadeepto Nag} \thanks{Indian Institute of Science Education and Research Pune, Pune-411008, Maharashtra, India, e-mail: suryadeepto.nag@students.iiserpune.ac.in}
\and \sc{Ananda Shikhara Bhat} \thanks{Indian Institute of Science Education and Research Pune, Pune-411008, Maharashtra, India, e-mail: ananda.bhat@students.iiserpune.ac.in}
\and \sc{Siddhartha P. Chakrabarty} \thanks{Department of Mathematics, Indian Institute of Technology Guwahati, Guwahati-781039, Assam, India, e-mail: pratim@iitg.ac.in}}
\date{}
\maketitle

\begin{abstract}
Chronic Myeloid Leukemia (CML) is a biphasic malignant clonal disorder that progresses, first with a chronic phase, where the cells have enhanced proliferation only, and then to a blast phase, where the cells have the ability of self-renewal. It is well-recognized that the Philadelphia chromosome (which contains the BCR-ABL fusion gene) is the ``hallmark of CML''. However, empirical studies have shown that the mere presence of BCR-ABL may not be a sufficient condition for the development of CML, and further modifications related to tumor suppressors may be necessary. Accordingly, we develop a three-mutation stochastic model of CML progression, with the three stages corresponding to the non-malignant cells with BCR-ABL presence, the malignant cells in the chronic phase and the malignant cells in the blast phase. We demonstrate that the model predictions agree with age incidence data from the United States. Finally, we develop a framework for the retrospective estimation of the time of onset of malignancy, from the time of detection of the cancer.
\end{abstract}

\section{Introduction}

Chronic Myeloid Leukemia (CML) is a malignant disorder of the hematopoetic stem cells (HSCs), which causes abnormal hematopoiesis, that results in increased survival and proliferation of blood cells. It was the first type of leukemia to be discovered \cite{goldman2010chronic} and accounts for 7-15\% of all adult leukemic cases \cite{cortes1996chronic}. When detected, a large fraction of patients are found to be asymptomatic, while others express symptoms such as fatigue, headache, muscle cramps, nausea, musculoskeletal pain, among others \cite{efficace2014patient}. In contrast to Acute Myeloid Leukemia (AML), CML is characterized by a period of gradual progression, which if left untreated for 3-5 years may lead to an accelerated proliferation of the cancer. Over the years, due to an improvement in medical knowledge and healthcare systems, the survival rates of CML patients have improved, with the cumulative survival rates increasing several folds from the corresponding numbers three decades ago \cite{chen2013trends}. However, to date, a large number of CML cases prove fatal and there remains much to learn about the dynamics of progression of CML.

The diagnosis of CML is associated with the detection of the Philadelphia (Ph) chromosome, described by a t(9;22) translocation that results in the fusion of the breakpoint cluster region gene (BCR) on chromosome 22 to the Abelson leukemia virus (ABL) gene on chromosome 9 \cite{sawyers1999chronic}. The formation of the resultant BCR-ABL oncogene and the activation of the BCR-ABL tyrosine kinase have been accredited as the causative factors of CML. Due to the expression of BCR-ABL tyrosine kinase, leukemic cells are conferred greater survival and proliferation abilities as compared to non-malignant blood cells \cite{jamieson2004chronic}. Although the BCR-ABL fusion gene and the production of tyrosine kinase are known to be the hallmarks of CML, empirical studies show that BCR-ABL can also be found in the blood of healthy individuals \cite{biernaux1995detection,bose1998presence,boquett2013analysis,ismail2014incidence}. This suggests that hematopoietic stem cells can mutate to a form that has the BCR-ABL fusion gene without immediately turning malignant. Moreover, the malignancy of CML is also attributed to mutations or deletions of tumor suppressors such as $p16$ and $p53$ \cite{ahuja1991spectrum,sawyers1999chronic}, apart from the sole presence of the BCR-ABL fusion gene.

While the literature on mathematical models of CML progression is fairly extensive \cite{fokas1991mathematical,neiman2000mathematical,michor2005dynamics,michor2005dynamics,horn2008mathematical}, most existing models do not consider the possibility of having healthy Ph+ individuals. To the best of our knowledge, there is only one existing model that accounts for non-malignant mutations that lead to BCR-ABL formation \cite{lecca2016accurate}. In their study, Lecca and Sorio \cite{lecca2016accurate} build two models of CML progression in order to explain the age incidence of CML. They build on previous work on two-mutation models \cite{armitage1957two, haeno2007evolution} and develop two variants of a model of CML where the first mutation leads to the formation of the BCR-ABL fusion gene and the second mutation leads to the deactivation of tumor suppressors, resulting in the onset of CML. They subsequently use their models to explain the age-incidence of CML in the United States of America. The age-incidence of CML has been explained in the past by another article \cite{michor2006age}, using a stochastic model based on the Moran process. While both these papers do well to explain the age incidence curves of CML, it is important to note that both models assume a pool of fixed size of cells, from where the initial mutations arise. In the paradigm of CML arising from stem-cell mutations, this assumption need not hold true if the size of the compartment of stem cells increases with age \cite{catlin2011replication}. Furthermore, Michor \textit{et al.} \cite{michor2006age} build a model where there is only one type of mutant cell, while Lecca and Sorio \cite{lecca2016accurate} work with two mutant cells in their model (while still considering only one type of leukemic cells). However, CML is neither clinically nor biologically thought of as a monophasic disease. Clinically, CML progressions are observed to have three phases, namely, a chronic phase, an accelerated phase, and a blast phase. These definitions are based on the amount of leukemic cells found in the blood of the patient, and not on distinct types of leukemic cells. It has also been argued that CML is fundamentally a biphasic disease in which activation of the $\beta$-catenin signalling pathway grants self-renewing abilities to differentiated leukemic cells, leading to the blast phase \cite{houshmand2019chronic}. Nevertheless, both the models \cite{michor2006age,lecca2016accurate}, despite considering monophasic compartments of leukemic cells, are extremely useful in accurately predicting the age-incidence of CML. This may be because an overwhelming majority ($\approx$~85\%) of CML is detected in the chronic phase \cite{faderl1999chronic}, implying that CML is predominantly monophasic at the time of detection. But outside the context of age at detection, it is worthwhile to investigate the age at which CML transitions from the chronic phase to the blast phase, and for this, a subsequent mutation(s) needs to be incorporated into models.

In this paper, we develop a three-stage stochastic model of CML progression from genesis to the blast crisis. We assume a growing compartment of HSCs, where a possible mutation leads to the formation of the BCR-ABL fusion gene. Subsequently, we leverage the observation that BCR-ABL is detected in a significant fraction of healthy individuals and incorporate a compartment of non-malignant stem cells (NMSCs) with BCR-ABL, which is assumed to have originated due to a mutation to a HSC. We further assume that a second mutation to these NMSCs results in the deactivation of tumor suppression and leads to the origin of malignant cells containing BCR-ABL, as has been previously hypothesized \cite{lecca2016accurate}. These malignant cells then lead to the proliferation of CML. Taking cues from the empirical literature \cite{jamieson2004chronic,houshmand2019chronic}, our model incorporates a third mutation that defines the transition from the chronic stage to the blast stage. We assume that leukemic cells in the blast stage can acquire a mutation that leads to gain of self-renewing abilities, which in turn causes them to have a very small death rate. The description of the proposed model is presented in Section \ref{Sec-Model}. Based on the stochastic model, and the consequently derived results for the age distributions of CML onset and detection, we develop a Bayesian framework that allows for retrospective inferences about the time of mutation and onset of CML based on the age of cancer detection and mass of cancer detected in Sections \ref{Sec-Bayesian} and \ref{Sec-Params}.

\section{The Model}
\label{Sec-Model}

\begin{figure}[h]
\centering
\includegraphics[width=0.8\linewidth]{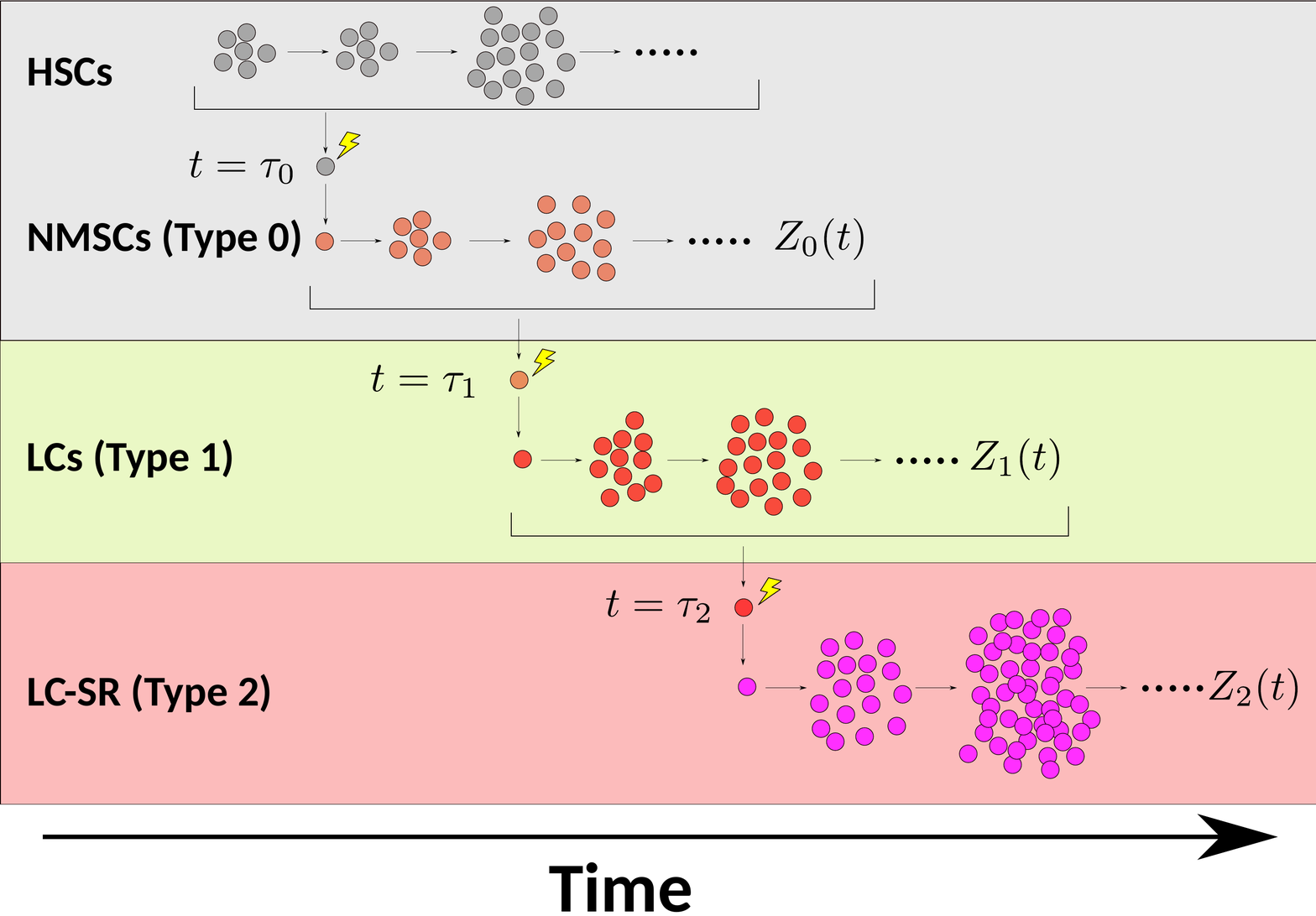}
\caption{Illustration of our three stage model for chronic myeloid leukemia (CML). Hematopoietic stem cells (HSC) mutate to non-malignant mutated stem cells (NMSCs), which lead to normal hematopoiesis (grey). NMSCs mutate to form leukemic cells (LC) that cause chronic phase CML (light green) and LCs mutate to form self-renewing leukemic cells (LC-SR) which leads to blast phase CML (light red).}
\label{fig:my_label}
\end{figure}

In our three-mutation model of cancer progression, the first mutation leads to the formation of non-malignant mutated stem cells (NMSCs or Type 0 cells) from the hematopoietic stem cells (HSCs), which differ from normal stem cells in the presence of BCR-ABL. The second mutation results in the formation of malignant leukemic cells (LCs or Type 1 cells) in the chronic phase, that are capable of enhanced proliferation by the deactivation or deletion of tumor suppression Finally, the third mutation results in the formation of self-renewing leukemic cells (LC-SRs or Type 2 cells), due to the activation of the $\beta$-catenin in the mature leukemic cells, which leads to the blast crisis phase (Figure \ref{fig:my_label}).  We model the growth of each of these types of cells as a branching process \cite{durrett2015branching} with a probability of mutating to the next stage (except for Type 2 cells). Although the actual dynamics of the progression is more complex than a simple three-stage system, and involves multiple differentiation stages of leukemic cells and possibly variations in birth and death rates of the different types of cells \cite{michor2005dynamics}, our model captures the essential features of the disease as the distinctions between different types of cells based on BCR-ABL and $\beta$-catenin activation may lead to more pronounced differences in progression dynamics than distinctions in different differentiation stages. The ability of models to capture the important features can be seen in  \cite{michor2006age} and \cite{lecca2016accurate}, both of which explain the age-incidence data well despite not considering multiple differentiation stages. 

\subsection{Type 0 Cells: Non-malignant Mutations}

\label{ss-Type0}
Existing models \cite{michor2006age,lecca2016accurate} have considered a compartment of constant size ($N=2000$) for the cells which give rise to mutations. However, since CML is thought to be caused by mutations to HSCs, and since the number of stem cells in an individual need not be constant from birth up to adulthood, we relax this assumption and consider a stem cell compartment that changes with age.  Motivated by literature on multiplication of HSCs upto adulthood \cite{catlin2011replication}, and following previous models of organismal growth \cite{erten2020zygote,de2022scaling}, we choose the growth function based on the assumption that the growth of hematopoietic stems cells to their carrying capacity will be logistic, wherein HSCs exist at birth and their population grows until they reach a maximum viable population. Accordingly, the logistic growth model of HSCs from birth is given by the differential equation,
\[\frac{dH}{dt}=rH\left(1-\frac{H}{J}\right),\]
where $H$ is the number of HSCs, $J$ is the carrying capacity of HSCs, $r$ is the growth rate of the number of cells and $t$ is time in years. Now suppose that each cell divides $\phi$ times per unit time. Therefore the number of cell divisions per unit time is given by $S=\phi H$. Similarly, the carrying capacity $K$ of the number of cell divisions per unit time is given by $K=\phi J$. Therefore, we can correspondingly write the differential equation for the number of cell divisions, which is given by,
\[\frac{dS}{dt}=rS\left(1-\frac{S}{K}\right).\]
Therefore, the number of cell divisions per unit time required to grow and maintain the compartment of HSCs (by adding new cells in the growth phase and replacing dead cells in the mature phase) is also logistic, scaled only by a factor that quantifies the rate of cell divisions (of each cell) per unit time $(\phi)$. The solution of this differential equation for the initial condition $S(0)=S_{0}$ is given by,
\[S(t)=\frac{K}{1+\frac{K-S_{0}}{S_{0}}e^{-rt}}.\]
Since mutations typically arise during cell division, we assume that the probability of a mutation being conferred in an individual in an interval is proportional to the number of cell divisions that take place in that interval. Specifically, we assume that the probability density $\displaystyle{f_{\tau_{0}}(t)}$ for $\tau_{0}$, the time at which the first Type 0 cell is formed, is given by,
\begin{equation}
\label{eq-logistic}
f_{\tau_{0}}(t)=\frac{\alpha K}{1+\frac{K-S_{0}}{S_{0}}e^{-rt}}.
\end{equation}
Here, $K$ is the number of cell divisions of HSCs per unit time in a fully grown human being (the carrying capacity of cell divisions), $S_{0}$ is the number of cell divisions of HSCs per unit time at birth, and $\alpha$ is a positive constant.

We model the stem cell growth after initial formation as a branching process. Let $Z_{0}(t)$ denote the number of Type 0 cells in the body, with a birth rate $a_{0}$ and death rate $b_{0}$, which results in a net growth rate of $\lambda_{0}:=a_{0}-b_{0}$. Note that a branching process implies that the line of NMSCs may go extinct before the formation of a Type 1 cell with some probability $\rho<1$. However, by scaling $\alpha$ by a constant (the inverse of the probability of going extinct within a suitably large time), the new $f_{\tau_{0}}(t)$ can be thought of as a probability density function for a mutation that does not go extinct in any realistic time period. Thus by choosing appropriate values for mutation rates, we can arrive at density functions, conditioned on non-extinction of the mutation (over some time period of interest). Henceforth, we shall work assuming that the parameter values have been chosen such that we are only considering cases where the mutation does not die out. We shall informally refer to these as ``permanent'' lineages of cells. Stem cells, however, are quiescent and few in number when compared to differentiated blood cells (both normal and leukemic). Furthermore, despite having the BCR-ABL fusion gene, these cells are non-malignant and do not have a particularly high rate of proliferation. This corresponds to low values of $a_0$ (of the order of normal HSCs). These NMSCs may further mutate to form malignant leukemic stem cells which will give rise to differentiated leukemic cells.
\begin{table}[h]
\centering
\begin{tabular}{cccc}
\hline
Type & Definition & Properties & Hematopoiesis  \\
\hline 
Type 0 & Non-Malignant Mutated Stem Cell & BCR-ABL production & Normal\\
\hline
Type 1 & Leukemic Cell & BCR-ABL production, enhanced  & CML Chronic Phase \\
&&survival and proliferation&\\
\hline
Type 2 & Self Renewing Leukemic Cell& BCR-ABL production, enhanced & CML Blast Crisis\\
&&survival and proliferation, &\\
&&$\beta$-catenin pathway, self-renewal&\\
\hline
\end{tabular}
\caption{Types of Mutant Cells}
\label{tab:Definitions}
\end{table}

\subsection{Type 1 Cells: Chronic Phase CML}
\label{ss-Type1}

\begin{figure}[h]
\begin{subfigure}{.5\textwidth}
\centering
\includegraphics[width=1\linewidth]{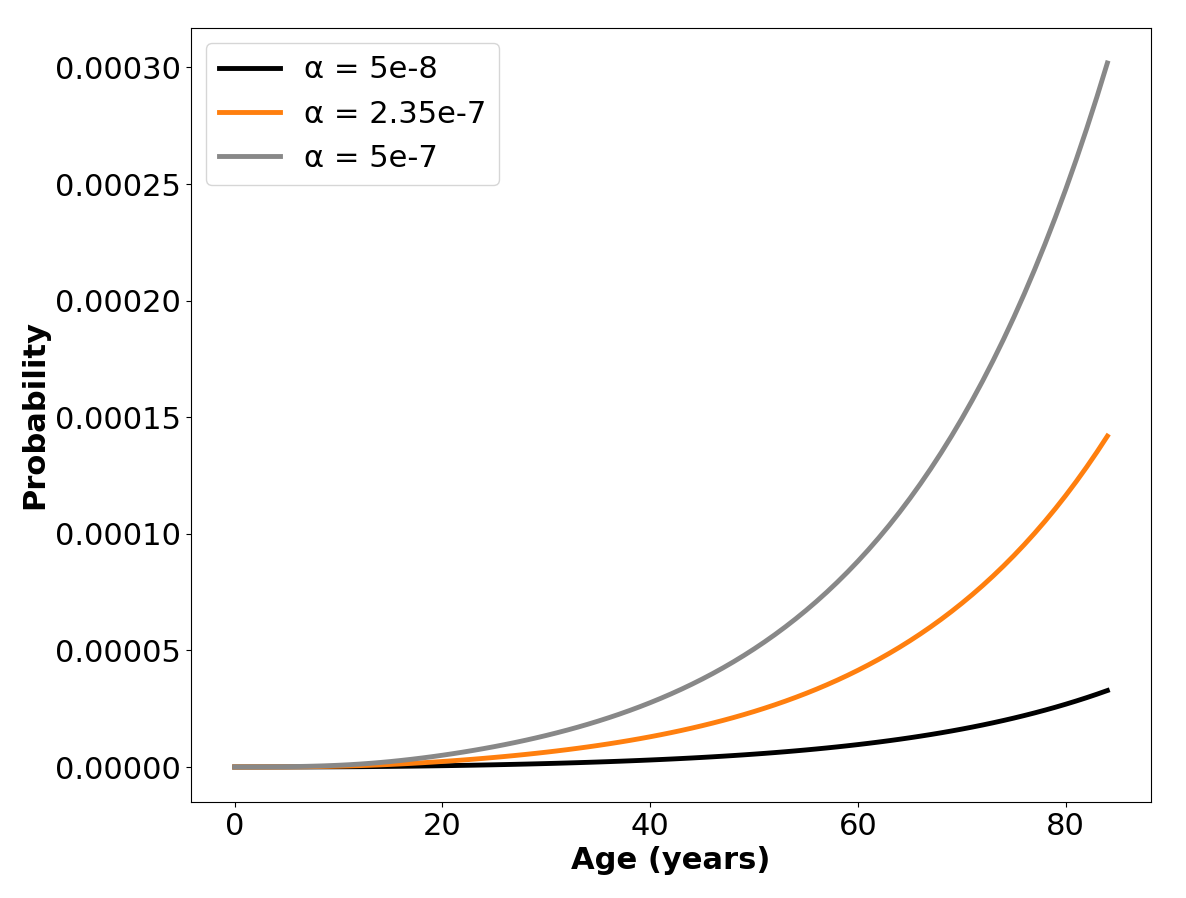}
\caption{}
\label{fig:First-T1-alpha-varied}
\end{subfigure}
\begin{subfigure}{.5\textwidth}
\centering
\includegraphics[width=1\linewidth]{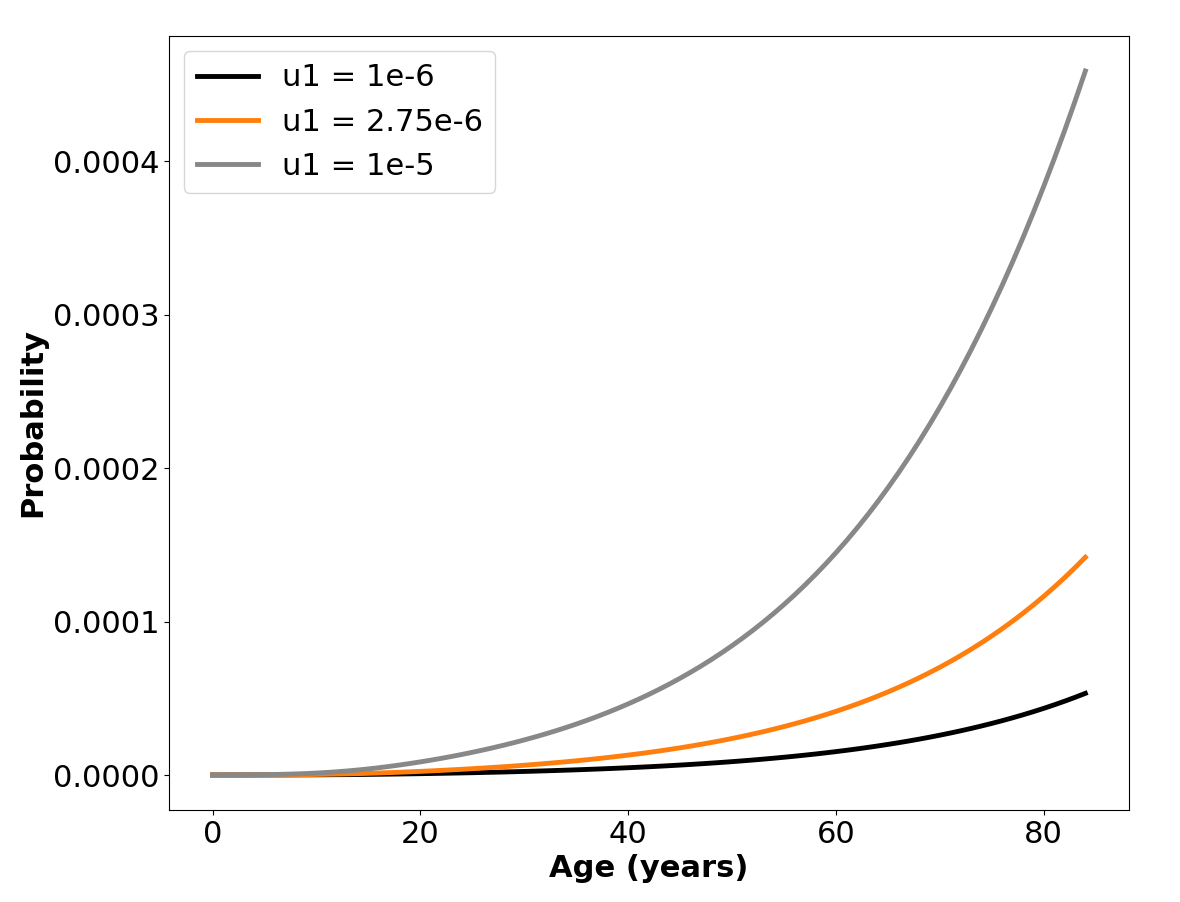}
\caption{}
\label{fig:First-T1-u1-varied}
\end{subfigure}
\caption{Age of formation of first Type 1 cell. We show the model predictions of the variation of the probability of acquiring a malignant leukemic cell with age. The parameter values are $K=14300$, $S_0=390$, $r=0.32$, $a_0=1.05$, $\lambda_0=0.05$, $a_1=4.75$, $\lambda_1=2$ (a) with three values considered $\alpha=5\times 10^{-8}$, $\alpha=2.35\times 10^{-7}$ and $\alpha=5\times 10^{-7}$ with $u_1$ kept constant at $2.75\times 10^{-6}$, and (b) with $u_1=1\times 10^{-6}$, $u_1=2.75\times 10^{-6}$ and $u_1=1\times 10^{-5}$ with fixed $\alpha=2.35\times10^{-7}$.}
\label{fig:First-T1} 
\end{figure}

Since BCR-ABL has been found to be present in a significant number of healthy individuals, following the hypothesis outlined in \cite{lecca2016accurate}, we assume that a second mutation (to NMSCs) is required to deactivate tumor suppressors, which leads to the development of LCs. Let $u_{1}$ be the rate of generation of a mutation that forms the first cell of a permanent lineage of Type 1 cells. The conditional CDF for the age of the patient at which the first Type 1 leukemic cell is formed is then given by \cite{durrett2015branching}, \[\mathbb{P}\left(\tau_{1}<t|\tau_{0}=t_{0}\right)=1-\left(1+\left(a_{0}/\lambda_{0}^{2}\right)u_{1}e^{\lambda_{0}\left(t-t_0\right)}\right)^{-1},\]
where we have conditioned on $\tau_{0}$, the age of the patient when the first Type 0 cell is formed. Taking the derivative with respect to $t_{1}$, we obtain the conditional probability density function,
\begin{equation}
\label{Eq-First-Type1}
f_{\tau_{1}|t_{0}}(t,t_{0})=\frac{a_{0}}{\lambda_{0}}u_{1}e^{\lambda_{0} \left(t-t_{0}\right)}\left(1+\left(a_{0}/\lambda_{0}^{2}\right)u_{1}e^{\lambda_{0}\left(t-t_{0}\right)}\right)^{-2}.
\end{equation}
To obtain the probability density function for the age of the patient at which the first malignant leukemic cell (that doesn't go extinct) is created, we note that,
\[f_{\tau_1}(t)= \int\limits_{0}^{t}f_{\tau_{1}|t_{0}}(t,t_{0})f_{\tau_{0}}(t_{0})dt_{0}
=\int\limits_{0}^{t}\frac{a_{0}}{\lambda_{0}}u_{1}e^{\lambda_{0}\left(t-t_{0}\right)}\left(1+\left(a_{0}/\lambda_{0}^{2}\right)u_{1}e^{\lambda_{0}\left(t-t_{0}\right)}\right)^{-2}\frac{\alpha K}{1+\frac{K-S_{0}}{S_{0}}e^{-rt_{0}}}dt_{0}.\]
This integral cannot be evaluated analytically and the result evaluated numerically is displayed in Figure \ref{fig:First-T1}.\\

\begin{figure}[H]
\centering
\includegraphics[width=0.5\linewidth]{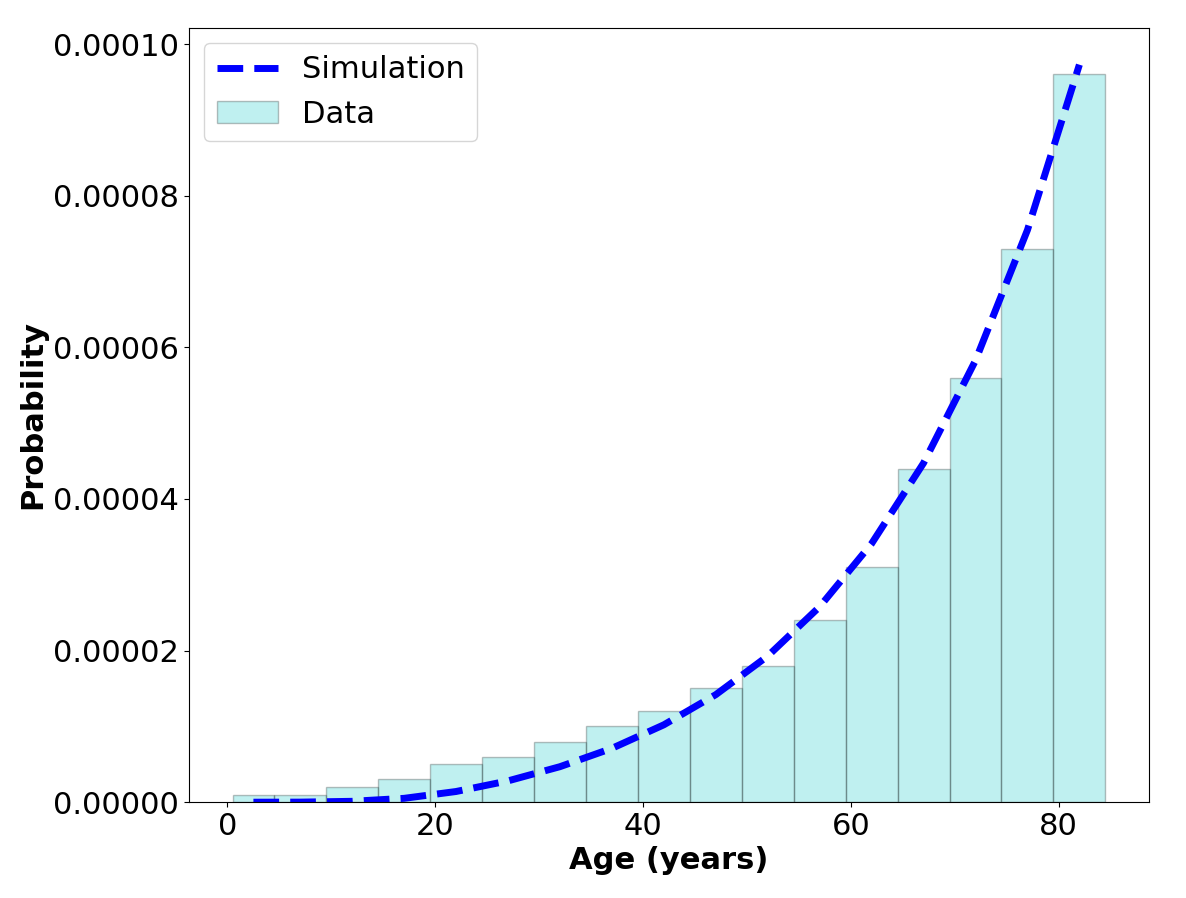}
\caption{The age incidence of CML. We compare the numerical result computed from the model with real data of CML incidence in the United States from 2007-2011. The model prediction shows a close correspondence with the data. The parameter values are $K=14300$, $S_0=390$, $r=0.32$, $\alpha=2.35\times 10^{-7}$, $a_0=1.05$, $\lambda_0=0.05$, $u_1=2.75\times 10^{-6}$, $a_1=4.75$, $\lambda_1=2$ and $M_1=10^5$.}
\label{fig:Age-Incidence}
\end{figure}

The median of the limiting distribution of the time between the first and the second mutation is given by \cite{durrett2015branching}
\begin{equation}
\label{eq-median-Type1}
t^{1}_{1/2}\approx \frac{1}{\lambda_{0}}\ln\left(\frac{\lambda_{0}^{2}}{a_{0}u_{1}}\right).
\end{equation}
Thus, for parameter values where $\lambda_{0}^{2}>a_{0}u_{1}e$, the median time of the mutation will be of the order of at least $\displaystyle{\frac{1}{\lambda_{0}}}$. Since the rate of growth of stem cells is relatively low, the median time will be large enough to make the probability of multiple mutations very low. Empirical literature \cite{biernaux1995detection,bose1998presence} also suggests that two mutations of type 0 cells to type 1 cells in a single individual is unlikely. It is therefore a feasible approximation to study the proliferation of Type 1 cells that arise from a single non-extinct mutation. In our model, Type 1 cells themselves follow a branching process with birth rate $a_{1}$, death rate $b_{1}$, and growth rate $\lambda_{1}:=a_{1}-b_{1}$. Thus, the (conditionally) expected number of cells at time $t$, namely $\mathbb{E}[Z_{1}(t)|Z_{1}(\tau_{1})]$ satisfies the ordinary differential equation,
\begin{equation}
\label{eq_ODE}
\frac{d}{dt}\mathbb{E}[Z_{1}](t)=\lambda_{1}\mathbb{E}[Z_{1}](t-\tau_{1}),
\end{equation} 
where we have dropped the conditioning term $Z_{1}(\tau_{1})$ for notational ease. By definition, $\tau_{1}$ is the time when the first Type 1 cell was formed, and thus, $Z_{1}(\tau_{1})=1$. Solving equation \eqref{eq_ODE} with this initial condition yields,
\[\mathbb{E}[Z_{1}](t)=\begin{cases}
0,&t<\tau_1\\
e^{\lambda_{1}\left(t-\tau_{1}\right)},&t\ge\tau_{1}
\end{cases}.\]
Under the above-mentioned assumption that the probability of multiple mutations that give rise to Type 1 cells occurring in the same individual is negligible, this expression serves as an approximation for the growth of Type 1 cells. Suppose then that the leukemia is detected when there are $M_{1}$ number of Type 1 cells \textit{i.e.,} $Z_{1}(t)=M_{1}$. Conditioned on the time of origin ($\tau_1$) of the first permanent Type 1 cell, the density function for the time of detection is given by the Gumbel distribution \cite{durrett2015branching},
\[f_{T_{M_{1}}|t_{1}}(t,t_{1})=\exp\left(-\frac{\lambda_{1}}{a_{1}}M_{1} e^{-\lambda_{1}\left(t-t_{1}\right)}\right) \frac{M_{1}\lambda_{1}^{2}}{a_{1}}e^{-\lambda_{1}\left(t-t_{1}\right)}.\]
Combining with equation \eqref{Eq-First-Type1} and using the previous approach of obtaining the marginal density by integrating, we observe that,
\[f_{T_{M_{1}}}(t)=\int\limits_{0}^{t}f_{T_{M_{1}}|t_{1}}(t,t_{1})f_{\tau_{1}}(t)dt_{1}=\int\limits_{0}^{t}\int\limits_{0}^{t_{1}}f_{T_{M_{1}}|t_{1}}(t,t_{1})f_{\tau_{1}|t_{0}}(t_{1},t_{0})f_{\tau_{0}}(t_{0})dt_{0}dt_{1}.\]
Substituting the functional forms of $\displaystyle{f_{T_{M_{1}}|\tau_{1}}(t_{1})}$, $\displaystyle{f_{\tau_{1}|\tau_{0}}(t_{0})}$, and $f_{\tau_{0}}(t_{0})$, we thus obtain,
\begin{equation}
\label{Eq-TM1-density}
f_{T_{M_1}}(t)=\int\limits_{0}^t\int_{0}^{t_{1}}\exp\left(-\frac{\lambda_{1}}{a_{1}}M_{1} e^{-\lambda_{1}\left(t-t_{1}\right)}\right)\frac{M_{1}\lambda_{1}^{2}}{a_{1}}e^{-\lambda_{1}\left(t-t_{1}\right)}\frac{a_{0}}{\lambda_{0}}\frac{u_{1}e^{\lambda_{0}\left(t_{1}-t_{0}\right)}}{\left(1+\left(a_{0}/\lambda_{0}^{2}\right)u_{1}e^{\lambda_{0}\left(t_{1}-t_{0}\right)}\right)^{2}}\frac{\alpha K}{1+\frac{K-S_{0}}{S_{0}}e^{-rt_{0}}}dt_{0}dt_{1}.
\end{equation}
We observe that for specific, reasonable parameter values, the model prediction for the age-incidence of CML agrees with clinical incidence data from United States (Table \ref{tab:age_inc_data}), as can be seen in Figure \ref{fig:Age-Incidence}. The data has been compiled from  Surveillance, Epidemiology, and End Results (SEER) Cancer Statistics Review \cite{seerdata}. 
\begin{table}[h]
\centering
\begin{tabular}{cc}
\hline
Age at Diagnosis &  Incidence Rate (per 100,000 individuals)\\
\hline
1-4 & 0.1\\\hline
5-9	& 0.1\\\hline
10-14 &	0.2\\\hline
15-19 &	0.3\\\hline
20-24 &	0.5\\\hline
25-29 &	0.6\\\hline
30-34 &	0.8\\\hline
35-39 &	1\\\hline
40-44 &	1.2\\\hline
45-49 &	1.5\\\hline
50-54 &	1.8\\\hline
55-59 &	2.4\\\hline
60-64 &	3.1\\\hline
65-69 &	4.4\\\hline
70-74 &	5.6\\\hline
75-79 &	7.3\\\hline
80-84 &	9.6\\\hline
\end{tabular}
\caption{The age-specific incidence of CML in the United States for 2007-2011, age-adjusted to the 2000 US Std population, as available on SEER Cancer Statistics Review, Table 13.13.}
\label{tab:age_inc_data}
\end{table}

We do not fit the data to estimate parameters as the data is collected for all detections of CML while we make certain assumptions in the prediction of the simulated age-incidence curve. For instance, we assume that CML is detected as soon as a tumor of size $M_1$ is formed in an individual. However, in reality, CML is detected through routine blood tests or following the onset of symptoms, and there will thus likely be variations in the mass of the tumor at the time of detection. The data may also contain some patients who are already in the blast phase of CML. We deal with this latter possibility in the next section.

\subsection{Type 2 Cells: Blast Crisis}
\label{ss-Type2}

Chronic phase CML, if left untreated, usually progresses to the blast phase, also known as the blast crisis. The blast crisis is marked by the presence of leukemic cells (Type 2) which not only have enhanced proliferation and survival but also the property of cell renewal. In our model, we assume that these Type 2 cells are formed due to the mutation to Type 1 cells. Let $u_{2}$ be the mutation rate of Type 1 cells becoming Type 2 cells. As derived above for Type 1 cells, the probability density function for the formation of the first Type 2 cell, conditioned on the time of formation of the first permanent Type 1, cell is given by \cite{durrett2015branching},
\begin{equation}
\label{Eq-First-Type2}
f_{\tau_{2}|t_{1}}(t,t_{1})=\frac{a_{1}}{\lambda_{1}}u_{2}e^{\lambda_{1}\left(t-t_{1}\right)}\left(1+\left(a_{1}/\lambda_{1}^{2}\right)u_{2}e^{\lambda_{1}\left(t-t_{1}\right)}\right)^{-2}.
\end{equation}
Therefore, 
\begin{eqnarray*}
f_{\tau_{2}}(t)&=&\int\limits_{0}^{t}f_{\tau_{2}|t_{1}}(t_{1},t)f_{\tau_{1}}(t_{1})dt_{1}\\
&=&\int\limits_{0}^{t}\frac{a_{1}}{\lambda_{1}}\frac{u_{2}e^{\lambda_{1} \left(t-t_{1}\right)}}{\left(1+\left(a_{1}/\lambda_{1}^{2}\right)u_{2}e^{\lambda_{1} \left(t-t_{1}\right)}\right)^{2}}\int\limits_{0}^{t_{1}}\frac{a_{0}}{\lambda_{0}}\frac{u_{1}e^{\lambda_{0} \left(t_{1}-t_{0}\right)}}{\left(1+\left(a_{0}/\lambda_{0}^{2}\right)u_{1}e^{\lambda_{0}\left(t_{1}-t_{0}\right)}\right)^{2}}\frac{\alpha K}{\left(1+\frac{K-S_{0}}{S_{0}}e^{-rt_{0}}\right)}dt_{0}dt_{1}.
\end{eqnarray*}
\begin{figure}[h]
\begin{subfigure}{.5\textwidth}
\centering
\includegraphics[width=1\linewidth]{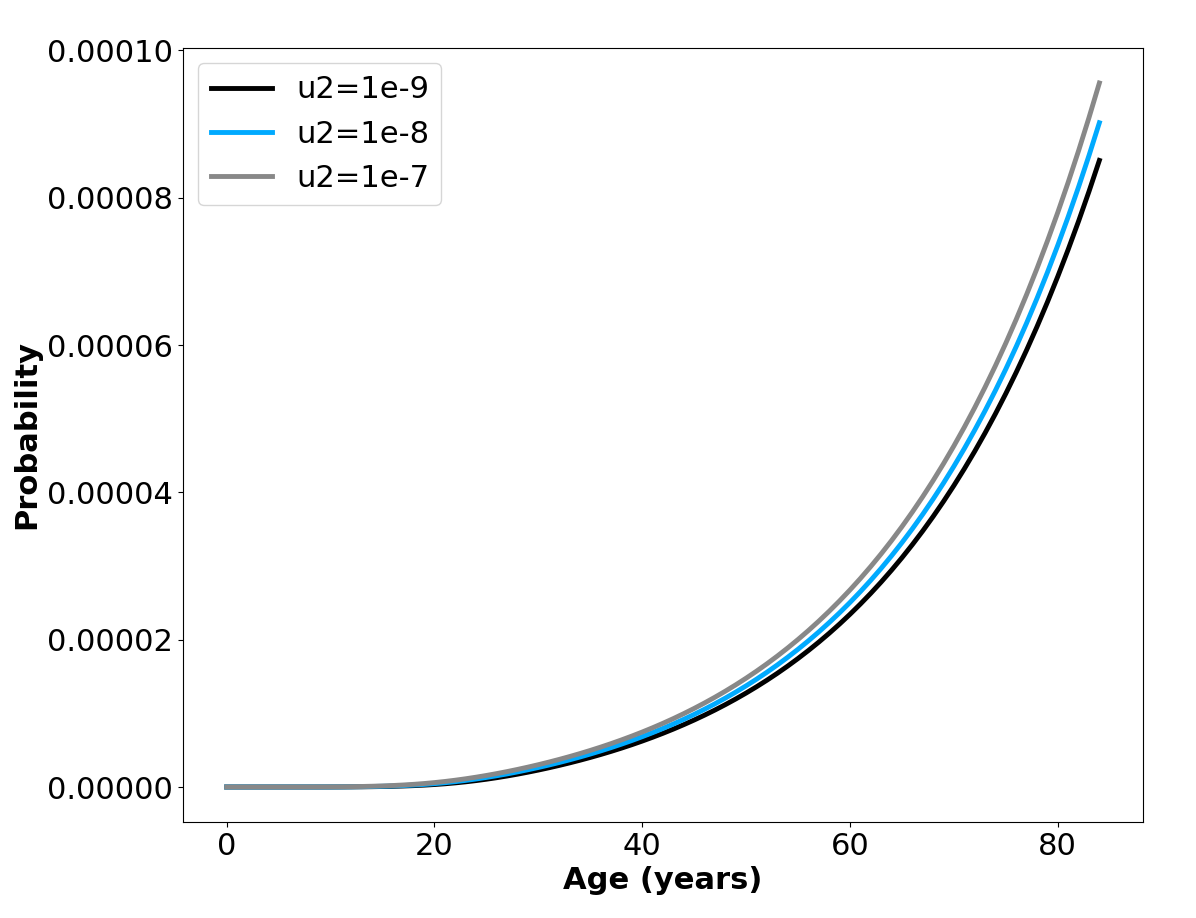}
\caption{}
\label{fig:First-T2-u2-varied}
\end{subfigure}
\begin{subfigure}{.5\textwidth}
\centering
\includegraphics[width=1\linewidth]{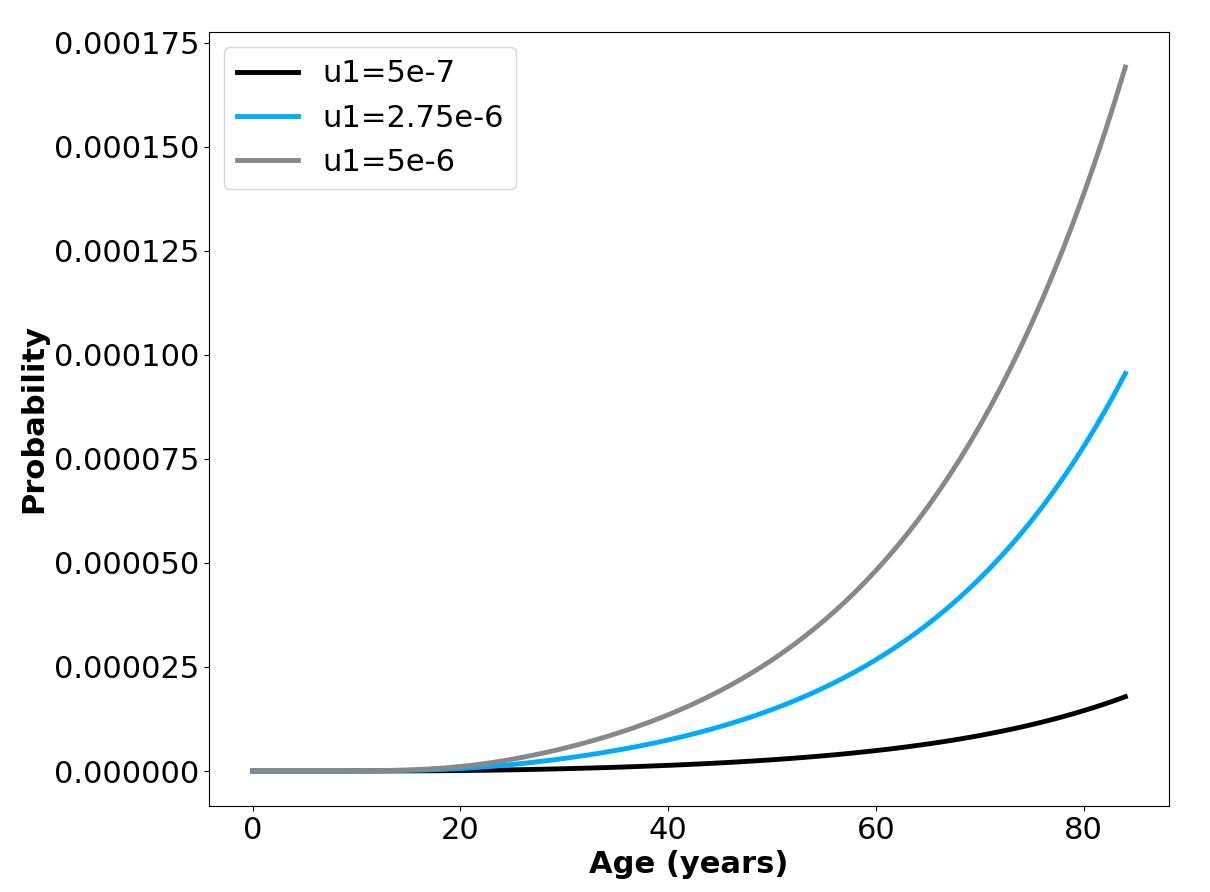}
\caption{}
\label{fig:First-T2-u1-varied}
\end{subfigure}
\caption{The Age-Incidence of Blast Phase CML. The age-dependence of the probability of time of first occurrence of self-renewing leukemic cells is shown. The parameter values are $K=14300$, $S_0=390$, $r=0.32$, $\alpha=2.35\times 10^{-7}$, $a_0=1.05$, $\lambda_0=0.05$, $a_1=4.75$, $\lambda_1=2$, and (a) fixed $u_1=2.75\times 10^{-6}$ with three values of $u_2$ considered, $u_2=10^{-9}$, $u_2=10^{-8}$ and $u_2=10^{-7}$ and (b) fixed $u_2=10^{-7}$ and three values of $u_1$ considered $u_1=5\times 10^{-7}$,$u_1=2.75\times 10^{-6}$ and $u_1=5\times 10^{-6}$}
\label{fig:First-Type2}
\end{figure}

The numerically integrated result of the distribution of the age of first formation of Type 2 cells is shown in Figure \ref{fig:First-Type2}. We can observe that the sensitivity of the the distribution to $u_2$ is low with entire orders of magnitude making a small difference (Figure \ref{fig:First-T2-u2-varied}). On the other hand, an order's difference in $u_{1}$ makes a significant difference to the distribution (Figure \ref{fig:First-T2-u1-varied}).

Given that most detections of CML happen during the chronic phase (as opposed to the accelerated or blast phases), equation \eqref{Eq-TM1-density} is the mathematical expression for the age-incidence of CML detection. However, in a small number of cases, it is possible that some leukemic cells have already acquired the ability of self-renewal by activating the $\beta$-catenin pathway. These Type 2 cells have high birth rates $a_{2}$ due to an enhanced expression of the BCR-ABL fusion protein and a negligible death rate $b_{2}\to 0$ due to the self renewal ability, leading to an abnormally large accumulation of leukemic cells in blood in a very short period. The property of self-renewal in these Type 3 cells implies the death rate $b_{2}$ of Type 2 cells is virtually 0. Hence, they follow a special type of branching process called the ``pure-birth process'' or the ``Yule process''. Thus, for Type 2 cells, $\lambda_{2}:=a_{2}-0=a_{2}$. This leads to the exponential growth of the leukemic cells which results in a blast crisis. It is also important to note that the assumption of a zero death rate implies that a Type 2 cell once formed will never go extinct and will necessarily lead to a blast crisis if left untreated.

At the time of detection, Type 2 cells (if present) are likely to be few in number, making it difficult to detect and isolate them immediately during diagnosis. It is therefore of value to study the probability of such cells already existing when the cancer is detected. Conditional to the founding of a permanent lineage of Type 1 cells, the probability that there already exists a mass of Type 2 cells at the time of detection is given by \cite{iwasa2006evolution,durrett2015branching},
\begin{equation}
\label{eq_Iwasa_type_2_exp}
\mathbb{P}(Z_{2}(T_{M_{1}})>0|\tau_{1})=1-e^{-\mu(M_{1},\tau_{1})},
\end{equation}
where,
\[\mu(M_{1},\tau_{1})=M_{1}u_{2}\lambda_{2}\int\limits_{\tau_{1}}^{\infty}\frac{e^{-\lambda_{1}(s-\tau_{1})}}{a_{2}-b_{2}e^{-\lambda_{2}(s-\tau_{1})}}ds.\]
Substituting  $a_{2}=\lambda_{2}$ and $b_{2}=0$, we obtain the simpler expression:
\[\mu(M_1,\tau_1)=M_1u_2\int_{\tau_1}^{\infty}e^{-\lambda_1(s-\tau_1)}ds=
\frac{M_1u_2}{\lambda_1}\biggl(1-0\biggr)=\frac{M_1u_2}{\lambda_1}.\]
We observe that this expression is independent of the time of diagnosis and the time of onset of leukemia (formation of the first Type 1 cell) and only depends on the mass of the cells detected, the growth rate of the type 1 cell, and the mutation rate of transitioning from Type 1 to Type 2 cells. Substituting this expression into equation \eqref{eq_Iwasa_type_2_exp}, we see that the probability of already having Type 2 cells, at detection, is given by, 
\[\mathbb{P}(Z_{2}(T_{M_{1}})>0)=1-\exp\left(-\frac{M_{1}u_{2}}{\lambda_{1}}\right).\]

\section{Retrospective Estimation of Age of Cancer Origin}
\label{Sec-Bayesian}

Due to the stochastic nature of cancer progression, it is difficult to estimate the duration of CML before it is detected. Therefore, it would be useful to be able to estimate the time of onset of the leukemia given that a mass of $M_{1}$ cells have been detected in a patient of age $T_{M_{1}}$, which depended on region specific parameter values. Furthermore, studies show that the latency period for children is much shorter than adults among patients of CML. Thus, it would also be useful to study how the time of onset or the latency period varies with the age of detection. For this purpose, we develop a framework to theoretically estimate the time of the formation of the first Type 0 and Type 1 cell, based on knowledge of parameter values and the mass and age of detection. To compute this, we use Bayes' theorem,
\[\mathbb{P}(A|B)=\frac{\mathbb{P}(B|A)\mathbb{P}(A)}{\mathbb{P}(A)}\]
to note that
\[f_{\tau_{1}|t_{M_{1}}}(t)=\frac{f_{t_{M_{1}}|\tau_{1}}(t)f_{\tau_{1}}(t)}{f_{T_{M_{1}}}(t_{M_{1}})}\]
is the equation for the probability that the first mutation that led to leukemia detected in a patient of age $t_{M_{1}}$ with $M_{1}$ leukemic cells arose when the patient was at age $t$. From this, we can find the mean of the distribution,
\[\mathbb{E}[\tau_{1}|t_{M_{1}}]=\int\limits_0^{t_{M_{1}}}tf_{\tau_{1}|t_{M_{1}}}(t)dt,\]
and the variance,
\[\text{Var}(\tau_{1}|t_{M_{1}})=\int\limits_0^{t_{M_{1}}}t^{2}f_{\tau_{1}|t_{M_{1}}}(t)dt-\left(\int\limits_{0}^{t_{M_{1}}}tf_{\tau_{1}|t_{M_{1}}}(t)dt\right)^2.\]
Similarly, for the time of formation of the first Type 0 cell, we have,
\[f_{\tau_{0}|t_{M_{1}}}(t)=\frac{f_{t_{M_{1}}|\tau_{0}}(t)f_{\tau_{0}}(t)}{f_{T_{M_{1}}}(t_{M_{1}})},\]
with the mean being,
\[\mathbb{E}[\tau_{0}|t_{M_{1}}]=\int\limits_{0}^{t_{M_{1}}}tf_{\tau_{0}|t_{M_{1}}}(t)dt,\]
and the variance being,
\[\text{Var}(\tau_{0}|t_{M_{1}})=\int\limits_{0}^{t_{M_{1}}}t^{2}f_{\tau_{0}|t_{M_{1}}}(t)dt-\left(\int\limits_{0}^{t_{M_{1}}}tf_{\tau_{0}|t_{M_{1}}}(t)dt\right)^{2},\]
where $f_{\tau_{0}|t_{M_{1}}}(t)$ is the probability density for the first Type 0 cell being formed at age $t$ given that a leukemia of mass $M_{1}$ has been detected in the patient at time $t_{m}$.
\begin{figure}[H]
\begin{subfigure}{.5\textwidth}
\centering
\includegraphics[width=1\linewidth]{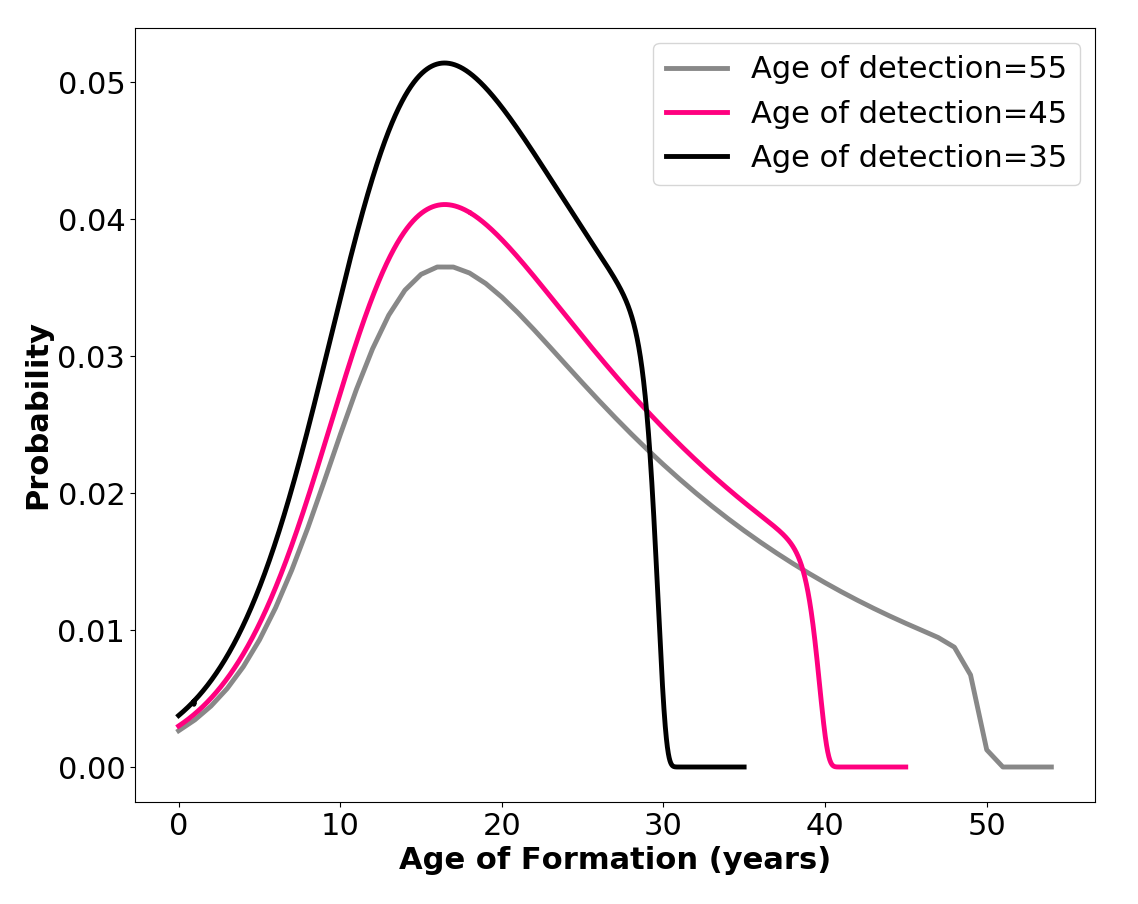}
\caption{}
\label{fig:Bayesian-Type0}
\end{subfigure}
\begin{subfigure}{.5\textwidth}
\centering
\includegraphics[width=1\linewidth]{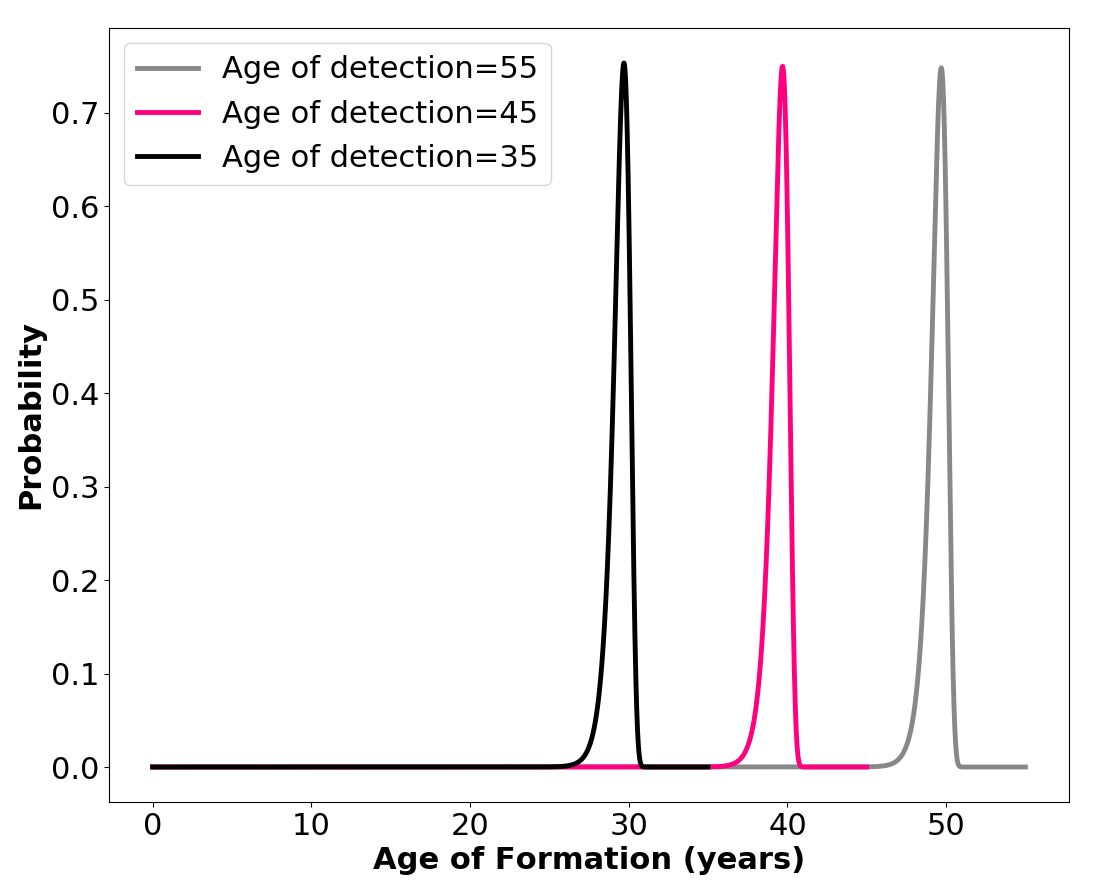}
\caption{}
\label{fig:Bayesian-Type1}
\end{subfigure}
\newline
\begin{subfigure}{1\textwidth}
\centering
\includegraphics[width=0.5\linewidth]{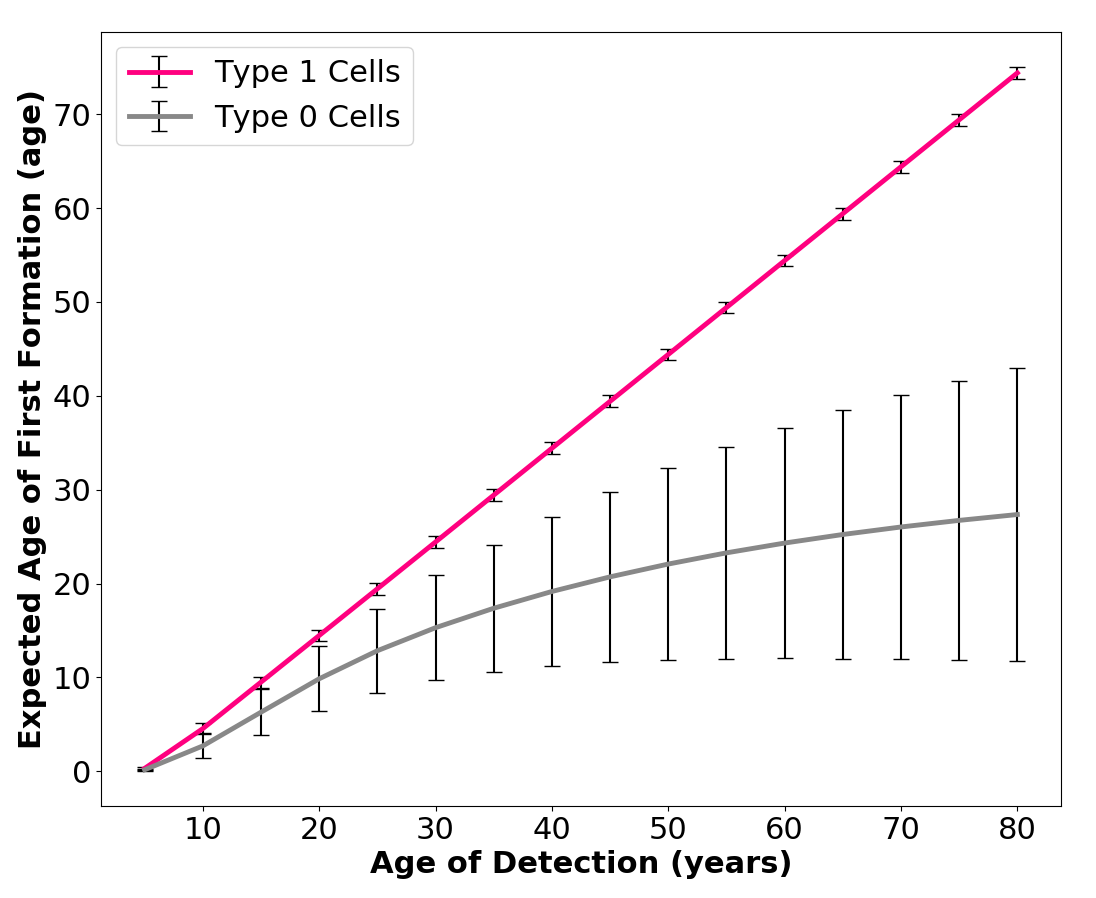}
\caption{}
\label{fig:Bayesian-Exp-Origin}
\end{subfigure}
\caption{Retrospective estimation of time of formation of mutated cells. We show the numerically calculated distributions for the age of formation of (a) the first Type 0 cell and (b) the first permanent Type 1 cell. We also show (c) how the expected time of formation varies with the time of detection. The error bars represent the standard deviation of the distributions. The parameter values are $K=14300$, $S_0=390$, $r=0.32$, $\alpha=2.35\times 10^{-7}$, $a_0=1.05$, $\lambda_0=0.05$, $u_1=2.75\times 10^{-6}$, $a_1=4.75$, $\lambda_1=2$ and $M_1=10^5$.}
\label{fig:Bayesian} 
\end{figure}

For the parameter values chosen, we find a higher variance in the expected time of formation of the first Type 0 cell (Figure \ref{fig:Bayesian-Type0}) than the corresponding quantity for Type 1 cells (Figure \ref{fig:Bayesian-Type1}). This difference can also be observed in the error bars in Figure \ref{fig:Bayesian-Exp-Origin}. This is because while stochasticity in the estimation of $\tau_{1}$ is only attributed to the growth dynamics of Type 1 cells, the stochasticity in the estimation of $\tau_{0}$ cells arises, not only from the growth dynamics of Type 2 cells, but also from the mass of Type 0 cells at the times when the first Type 1 cell is formed $(Z_{0}(\tau_{1}))$, which itself is a random variable. We also find the expected latency period ($T_{M_{1}}-\tau_{1}$) comes out to be approximately six years. Consistent with literature, we find that the latency period increases with age \cite{abecasis2020cancer}. The time from the first mutation to the time of detection also increases with age (Figure \ref{fig:Bayesian-Exp-Origin}). The average time between the formation of the first Type 0 cell and the time of detection comes out to be marginally over 25 years, which agrees with clinical data from survivors of the Chernobyl nuclear disaster \cite{ernst2020molecular}.

\section{Choice of Parameter Values}
\label{Sec-Params}

Some of our parameter values are based on literature, others are chosen such that they give realistic predictions that are consistent with empirical knowledge and model assumptions. Recent studies \cite{abkowitz2000vivo,catlin2011replication} find that the number of HSCs in adults is around 11000. Catlin \textit{et al.} \cite{catlin2011replication} also find that the optimal value for the number of HSCs at birth is 300 and the rate of replication is once in 40 weeks. Since we are interested in the number of cell divisions per year, we calculate $\displaystyle{K=11000\times \frac{52}{40}=14300}$ and $\displaystyle{S_{0}=300\times \frac{52}{40}=390}$. Following Catlin \textit{et al.} \cite{catlin2011replication}, we know that the mature size of the HSC compartment is reached (most likely) by the age of 18.5 years. Therefore, we set $r=0.32$ such that at least 90\% of the mature size of the stem cell compartment is reached by the age of 18.5 years. Since the probability density of the occurrence of a mutation is given by equation \eqref{eq-logistic}, we can find the total probability of having a mutation by integrating it. Assuming $t_{\max}$ years of age as the maximum, we have,
\[\mathbb{P}(Z_{0}\left(t_{\max}\right)>0)=\int\limits_{0}^{t_{\max}}\frac{\alpha K}{1+\frac{K-S_{0}}{S_{0}}e^{-rt}}dt=\frac{\alpha K}{r}\ln\left(\frac{S\left(e^{rt_{\max}}-1\right)+K}{K}\right).\]
We set $t_{\max}=85$ and choose $\alpha=2.35\times 10^{-7}$ such that the fraction of individuals who would have experienced the mutation in their lifetime is similar to the results reported by \cite{biernaux1995detection} and \cite{bose1998presence},  who find that roughly 30\% healthy individuals have BCR-ABL in their blood. For our choice of variables, this fraction comes out to be 29.82\%. This may lead to an underestimate as not all individuals considered in the empirical studies were over the age of 85. Regardless, since $\alpha$ influences the probability linearly, the parameter value chosen will be in the right order of magnitude.

The next set of parameters are $a_{0}$, $\lambda_{0}$, $u_{1}$, $a_{1}$, $\lambda_{1}$, $u_{2}$. Since these parameters are difficult to put a value on from literature, we choose them such that they lead to age incidence curves which are similar to empirical data, as presented in Figure \ref{fig:Age-Incidence}. However, we ensure that the order of magnitude of the parameter values are reasonable. Since stem cells replicate roughly once a year, the birth rate for Type 0 cells is set at $a_{0}=1.05$. Since these cells are not malignant, we choose a small value of $\lambda_{0}=0.05$. $u_{1}$ is chosen such that only a small fraction of individuals with Type 0 cells actually develop Type 1 cells. This is verified by studying the median time to the formation of the first Type 1 cell given by equation \ref{eq-median-Type1}. We find that for the chosen value of the parameters, $u_{1}=2.75\times 10^{-6}$, the median time to the formation of the first Type 1 cell is over 135 years after the formation of the first Type 0 cell. This implies that very few people will get CML. This also implies that the probability of two successive mutations in the same individual is extremely low. Therefore, our approximation that there is at most a single line of permanent Type 1 cells remains valid. Furthermore, the average time of detection is given by \cite{durrett2015branching} as,
\[\mathbb{E}[T_{M_{1}}-\tau_{1}]=\frac{1}{\lambda_1}\ln\left(\frac{M_{1}\lambda_{1}}{a_{1}}\right)-\frac{1}{\lambda_{1}}\int\limits_{0}^{\infty} e^{-x}\ln(x)dx.\]
For our parameter values, we find this quantity to be a little over 5 years, which is a reasonable result. For the calculation of $u_2$, we use the equation for the median time to the first Type 2 cell from the time of formation of the first Type 1 cell, given by,
\[t_{1/2}^{2}\approx\frac{1}{\lambda_{1}}\ln\left(\frac{\lambda_{1}^{2}}{a_{1}u_{2}}\right).\]
We have used parameter values between $u_{1}=10^{-7}$ and $u_{2}=10^{-9}$ which approximately yield median times to first type 2 cells between 8 years and 10 years, respectively. These correspond to approximately 3 years and 5 years after the expected time of detection, respectively, both of which agree with empirical knowledge. The choice of $M_{1}$, the abundance of leukemic cells at the time of detection in chronic phase is chosen as $10^5$, following the order of magnitude chosen in earlier studies \cite{michor2005dynamics}.

\section{Conclusion} 

The progression of cancer is a complex biological process that involves several stages with differences at the cellular level which may have ramifications at the level of the tissues, organs or a groups of organs. CML is simpler to model than other forms of cancer due to the fact that it is present in the blood, and can therefore be analyzed in a non-spatial context. In this article, we present a new stochastic model of CML that incorporates clinical observations and empirical knowledge. The definition and formalisms for the three stages defined in our model are based on three biological facts, namely, the presence of the BCR-ABL fusion gene in a significant fraction of healthy individuals, the property of enhanced survival and proliferation of leukemic cells conferred by the presence of BCR-ABL in malignancies in the chronic phase of the disease, and the additional activation of $\beta$-catenin pathway in cells that enables self-renewal in leukemic cells in the  blast phase of the disease. These stages also align with the clinical definitions of the phases of CML progression from healthy individuals to chronic phase CML to blast phase CML. Based on this, the model allows us to find the distribution of mutations and tumors in the human population, as a function of age and we use this to show significant parity between model predictions and clinical data of CML incidence in the United States. Working in a probabilistic framework also allows us to develop an approach to retrospectively estimate the time of BCR-ABL conferring and the time of onset of CML, based on the mass and time of detection, and knowledge of other biological parameters. Nevertheless, the model is simplistic in assuming homogeneity of cells within each stage, apart from differences induced by stochasticity. Future studies can improve our model by incorporating more biological subtleties such as differentiation stages, that distinguish between the several intermediate progenitor cells and fully differentiated leukemic cells, although this may make it difficult to model analytically and calls for more computation-based approaches.

\section*{Acknowledgement}

SPC was supported by Grant. No. MTR/2019/000225 from the Science and Engineering Research Board, India.

\section*{Declaration of Competing Interests}

The authors declare that they have no competing interests.

\bibliographystyle{apalike}
\bibliography{bibliography}

\end{document}